\documentclass{llncs}
\usepackage[utf8x]{inputenc}
\usepackage[english]{babel}
\usepackage{url}
\usepackage{amsmath, amssymb, amsfonts, amsbsy, bm}

\usepackage{mathtools}

\usepackage{graphicx}
\usepackage{stfloats}

\usepackage{caption, subcaption}
\usepackage{pgf, tikz}
\usetikzlibrary{patterns, shapes,arrows,fit,calc,positioning}
\tikzset{>=latex}

\definecolor{myblue}{RGB}{204,236,255}
\definecolor{mygreen}{RGB}{203,252,158}

\usepackage{mathtools}

\graphicspath{{pictures/}}
\DeclareGraphicsExtensions{.pdf,.png,.jpg,.eps}

\begin{document}

\tikzstyle{packet}=[draw, rectangle, fill=lightgray]

\title{On the Doubly Sparse Compressed Sensing Problem}

\titlerunning{Doubly}
\author{Grigory Kabatiansky \inst{1}, Serge
Vl\v{a}du\c{t} \inst{1,2}, Cedric Tavernier \inst{3} }
\institute{ Institute for Information Transmission Problems,
Russian Academy of Sciences, Moscow, Russia, \and Institut  de
Math\'ematiques  de Marseille, Aix-Marseille Universit\'e, IML,
Marseille,  France \and Assystem AEOS, France\\
\email{kaba@iitp.ru, vl@iitp.ru, tavernier.cedric@gmail.com}}
\thanks{The work of Grigory Kabatiansky and Serge
Vl\v{a}du\c{t} was carried out at the Institute for Information
Transmission Problems of the Russian Academy of Sciences at the
expense of the Russian Science Foundation, project
no.~14-50-00150.}

\maketitle              

\begin{abstract}
A new  variant of the  Compressed Sensing problem is
investigated when the number of measurements corrupted by
errors is upper bounded by some value $l$ but there are no more
restrictions on errors. We prove that in this case it is enough
to make $2(t+l)$ measurements, where $t$ is the sparsity of
original data. Moreover for this case a rather simple recovery
algorithm is proposed. An analog of the Singleton bound from
coding theory is derived what proves optimality of the
corresponding measurement matrices.
\end{abstract}

\section{Introduction and Definitions}
\label{intro}

A  vector $x=(x_{1},\ldots,x_{n})\in \mathbb{R}^{n}$ in
$n$-dimensional vector space $ \mathbb{R}^{n}$ called
$t$-sparse if its  Hamming weight $wt(x)$ or equivalently its
$l_{0}$ norm  $||x||_{0}$ is at most $t$, where by the
definition
 $wt(x)=||x||_{0}=|\{i: x_{i}\neq
0\}|$.  
Let us recall  that the Compressed Sensing (CS) Problem
\cite{Donoho,Tao} is   a
 problem of reconstructing of an $n$-dimensional  $t$-sparse
 vector $x$ by a few ($r$) linear measurements $s_{i}=\langle h^{(i)},x
 \rangle$ (i.e. inner product of vectors $x$ and $h^{(i)}$),
 assuming that measurements  $(h^{(i)},x)$ are known with some
 errors $e_{i}$, for $i=1,\ldots,r$. Saying in other words, one needs to construct an $r\times n$ matrix
 $H$ with minimal number of rows $h^{(1)},\ldots,h^{(r)}$,
 such that
 the following equation
 \begin{equation}
\hat{s}=Hx^{T}+e, \label{CS}
\end{equation}
has either a unique $t$-sparse  solution or all such solutions
are ``almost equal". The compressed sensing problem was mainly
investigated under the assumption that the vector
$e=(e_{1},\ldots,e_{r})$, is called the error vector, has
relatively small Euclidean norm (length) $||e||_{2}$. We
consider another problem's statement assuming that the error
vector $e$ is also sparse  but its Euclidean norm can be
arbitrary large. In other words, we consider the doubly sparse
CS problem when    $||x||_{0}\leq t$ and $||e||_{0}\leq l$.
 The assumption  $||e||_{0}\leq l$ was first time considered in
\cite{KabaVlad} as a proper replacement for discrete version of
CS-problem of usual assumption that an error vector $e$ has
relatively small Euclidean norm. 
\begin{definition}
 A  real  $r\times n$ matrix $H$ called a $(t,l)$-compressed sensing (CS)
matrix if
\begin{equation}
 ||Hx^{T}-Hy^{T}||_{0}\geq 2l+1  \label{L1}
\end{equation}
for any two distinct vectors $x,y\in \mathbb{R}^{n}$ such that
$||x||_{0}\leq t$ and $||y||_{0}\leq t$.
\end{definition}
This definition immediately leads (see \cite{KabaVlad}) to the
following

\begin{proposition}
 A real $r\times n$ matrix $H$ is  a $(t,l)$-CS
matrix iff
\begin{equation}
||Hz^{T}||_{0}\geq 2l+1  \label{L2}
\end{equation}
 for any nonzero vector $z\in \mathbb{R}^{n}$ such that $||z||_{0}\leq 2t$.
\end{proposition}
 Our main
result is an explicit and simple construction of $(t,l)$-CS
matrices with $r=2(t+l)$ for any $n$.  We show  this value of
$r$ is the minimal possible for $(t,l)$-CS matrices by proving
an analog of the well-known in coding theory Singleton bound
for the compressed sensing problem. Besides that we propose an
efficient recovery (decoding) algorithm for the considered
double sparse CS-problem.

\section{Optimal  Matrices for Doubly Sparse Compressed Sensing Problem}

 We start with constructing of $(t,l)$-CS matrices. Let a real  $\tilde{r}\times n$ matrix
 $\tilde{H}$ be a parity-check matrix of an $(n,n-\tilde{r})$-code code over $\mathbb{R}$, correcting $t$
 errors, i.e. any $2t$ columns $\tilde{h}_{i_{1}},\ldots,\tilde{h}_{i_{2t}}$ of  $\tilde{H}$ %
are linearly
 independent. And let $G$ be a  generator matrix
 of an $(r,\tilde{r})$-code  over $\mathbb{R}$ of length $r$, correcting $l$ errors.
 Let matrix $H$ consists of the columns $h_{1},\ldots, h_{n}$, where
 \begin{equation}
 h_{j}^{T}=\tilde{h}_{j}^{T}G  \label{concat}
\end{equation}
 and transposition $^{T}$ means, that  vectors $h_{j}$ and $\tilde{h}_{j}$ are
 considered in (4) as row vectors,   i.e.
 \begin{equation}
 H=G^{T}\tilde{H} \label{matrix}
\end{equation}
 In other words, we encode columns of parity-check matrix
 $\tilde{H}$, which is already capable to correct $t$ errors,  by a code,
 correcting $l$ errors, in order to restore correctly the syndrom
 of $\tilde{H}$.

 \begin{theorem}
\label{th:constr}  Matrix $H=G^{T}\tilde{H}$ is a $(t,l)$-CS
matrix.
\end{theorem}
\vspace{0.05in}\noindent \textit{Proof.}  According to
Proposition 2 it is enough to prove that $||Hz^{T}||_{0}\geq
2l+1$ for any nonzero vector $z\in \mathbb{R}^{n}$ such that
$||z||_{0}\leq 2t$. Indeed, $u=\tilde{H}z^{T}\neq 0$  since any
$2t$ columns of $\tilde{H}$
 are linear
 independent. Then $Hz^{T}=G^{T}\tilde{H}z^{T}=G^{T}u=(u^{T}G)^{T}$
 and $u^{T}G$ is a nonzero vector of a code over $ \mathbb{R}$, correcting $l$
 errors.
 Hence $||Hz^{T}||_{0}=||u^{T}G||_{0}\geq 2l+1$.
  \hfill $\square $\smallskip

  Now let us choose the well known Reed-Solomon (RS) codes (which are a particular case of
  evaluation codes construction) as both constituent codes. 
  The  length of the RS-code is restricted by the
  number of elements in the field so in the case of $\mathbb{R}$
   the length of evaluation code can be arbitrary large.
  Indeed, consider the corresponding evaluation code $\mathbb{RS}_{(n,k)}=\{(f(a_{1},\ldots,f(a_{n})): \deg f(x)<k\}$,
  where $a_{1},\ldots,a_{n}\in \mathbb{R}$ are
   $n$ different real numbers.
   The  distance of the $\mathbb{RS}_{(n,k)}$ code $d= n-k+1$ since the number of roots of a 
   polynomial cannot exceed its degree and hence $d\geq n-k+1$,but, on the other hand, the Singleton bound states that
   $d\leq n-k+1$ for any code, see \cite{MacW}. Therefore the
   resulting matrix $H$ is a $(t,l)$-CS
matrix with $r=2(t+l)$. The next result, which is a
generalization of the Singleton bound for the doubly sparse CS
problem, shows these matrices are optimal in the sense having
the minimal possible number $r$ of  linear measurements.

\begin{theorem}
\label{th:Sing}   For any  $(t,l)$-CS $r\times n$-matrix
\begin{equation}
 r\geq 2(t+ l).  \label{sing}
\end{equation}
\end{theorem}
\vspace{0.05in}\noindent \textit{Proof.}  Let $H$ be any
$(t,l)$-CS matrix of size $r\times n$, i.e.,
$||Hz^{T}||_{0}\geq 2l+1$ for any nonzero vector $z\in
\mathbb{R}^{n}$ such that $||z||_{0}\leq 2t$. And let
$H_{2t-1}$ be the $(2t-1)\times n$ matrix consisting of the
first $2t-1$ rows of $H$. There exists  a nonzero vector
$\hat{z}=(\hat{z}_{1},\ldots,\hat{z}_{2t},0,0,\ldots,0)\in
\mathbb{R}^{n}$ such that $H\hat{z}^{T}=0$ (a system of linear
homogenious equations with the number of unknown variables
larger than the number of equations has a nontrivial solution).
Then  $||H\hat{z}^{T}||_{0}\leq r-(2t-1)$ and finally $r\geq
2t+2l$ since $||H\hat{z}^{T}||_{0}\geq 2l+1$.
  \hfill $\square $\smallskip

  \section{Recovery Algorithm for Doubly Sparse Compressed Sensing Problem}

  Let us start from a simple remark that for  $e=0$ recovering of the original sparse vector $x$,
   i.e., solving the equation (1), is the same as syndrome decoding of some code (over $\mathbb{R}$)
 defined by matrix $H$  as a parity-check matrix.
 In general, syndrome $s=Hx^{T}$ is known with some error, namely, as
 $\hat{s}=s+e$ and therefore we additionally encoded  columns of $H$ by some error-correcting code
 in order to recover
 the original syndrome $s$ and then apply usual syndrome decoding algorithm.
 Therefore recovering, i.e., decoding
  algorithm for constructed in previous chapter optimal matrices is in some sense a
  ``concatenation" of  decoding algorithms of constituent codes.\\

   Namely, first we
  decode vector $\hat{s}=s+e$ by a decoding algorithm of the code
  with generator matrix $G$. Since $||e||_{0}\leq l$ this algorithm
  outputs the correct syndrome $s$. After that we form the syndrome $\tilde{s}$ by selecting first $\tilde{r}$
  coordinates of $s$
  and then apply syndrome decoding algorithm (of the first code with parity-check matrix $\tilde{H}$) for the following syndrom
  equation
  \begin{equation}
 \tilde{s}=\tilde{H}x^{T}.  \label{sindrom}
\end{equation}
Now let us discuss a right choice of constituent codes. It is
very convenient to use the class of Reed-Solomon codes over
$\mathbb{R}$. There are well known algorithms of their decoding
up to half of the code distance (bounded distance decoding, see
\cite{MacW}), for instance, Berlekamp-Massey algorithm, which
in our case (codes over $\mathbb{R}$) is known also as Trench
algorithm, see \cite{Trench}. Hence the total decoding
complexity does not exceed $O(n^{2})$  operations over real
numbers. Moreover we can even decode these codes over their
half  distances by application of Guruswami-Sudan list decoding
algorithm
\cite{Sudan}.\\

It is well known  that encoding-decoding procedures of
Reed-Solomon codes    become more simple in the case of cyclic
codes, when the set $a_{1},\ldots,a_{n}$ is a cyclic group
under multiplication. In order to do it let us consider
$a_{1},\ldots,a_{n}$ as complex roots of degree $n$ and define
our codes through their ``roots", i.e. our codes consist of
{\it polynomials} $f(x)$ over $\mathbb{R}$ such that $f(e^{2\pi
i \frac{m}{n}})=0$ for $m\in \{-s,\ldots,-1,0,+1,\ldots,+s\}$
with $s=t$ for the first constituent code and $s=l$ for the
second. It easy to check that such codes achieve the Singleton
bound with $d=2s+2$, so the corresponding doubly sparse code
has redundancy $r=2(t+l+1)$ what is slightly larger than the
corresponding Singleton bound, but in return these codes can be
decoded via FFT.

\section{Discussion - no small errors case  and
slightly beyond}

  Let us note that  the initial  papers on Compressed Sensing
  especially stated
   that this new technique (application of $l_1$ minimization instead of $l_0$)
   allows to recover information vector $x\in \mathbb{R}^n$
   in  case when not many coordinates of $x$ were affected by errors.
   For instance,  ``one can introduce errors of arbitrary large sizes and still
   recover the input vector exactly by solving a convenient linear
   program...", see in \cite{Tao-Codes}.
   To achieve such performance some special restriction on matrix $H$ was placed,
   called Restricted Isometry Property  (RIP), as follows
\begin{equation}
(1-\delta_{D}) ||x||_2\leq  ||Hx^{T}||_2\leq  (1+\delta_{D}) ||x||_2 ,  \label{RIP}
\end{equation}
for any vector $x\in \mathbb{R}^n : ||x||_0\leq D$, where
$0<\delta_D<1$.
 The smallest possible $\delta_D$ called the isometry constant. \\
 Then typical result in  \cite{Tao-Codes} (Th. 1.1 ) is of the following form \\
	{\it ``if $\delta_{3t}+3\delta_{4t}<2$ then the solution of linear programming problem is unique and equal to $x$"}\\
Let us note that the condition $\delta_{3t}+3\delta_{4t}<2$
implies $\delta_{4t}<2/3$ (of course, it implies that
$\delta_{4t}<1/2$, but for us enough to have $\delta_{4t}<1$).
Hence $Hx^{T}\neq \bf 0$ for any nonzero $x$ with $wt(x)\leq
4t$, or in other words, an error-correcting  code (over reals)
corresponding to such parity-check matrix $H$ has the minimal
distance at least $4t+1$ and  can correct $2t$ errors (instead
of $t$). So  we lost twice in error-correction capability but
maybe linear programming provides more easier way for decoding
? In fact, NOT, since it is well known in coding theory that
such problem can be solved rather easily (in complexity) over
{\it any infinite field} by usage of the corresponding
Reed-Solomon codes and known decoding algorithms. In case of
real number or complex number fields one can use just an RS
code with Fourier parity-check matrix, namely,
$h_{j,p}=exp{(2\pi i \frac{jp}{n})}$, where for complex numbers
$p\in \{1,2,...,n\},\; j=a,a+d,a+2d,\ldots,a+(r-1)d$,
and ``reversible" RS-matrix H for real numbers, where  $p\in \{1,2,...,n\},\; j\in \{-f,-f+1,\ldots,0,1,\ldots, f\}$
and $r=2f+1$.\\
Fortunately, matrices with the RIP property allow to correct
not only
 sparse errors but also additional errors with arbitrary
support but relatively small (up to $\varepsilon$) Euclidean
norm. Again, the RIP property is good for linear programming
decoding but is too strong in general. Namely,  it is enough to
have the following property
\begin{equation}
\lambda_{2t} ||x||_2\leq  ||Hx^{T}||_2,  \label{RIPweak}
\end{equation}
for any vector $x\in \mathbb{R}^n : ||x||_0\leq 2t$, where
$\lambda_{2t}>0$ and the largest such value we call {\it
extension constant}. Indeed, then for any two solutions $x$ and
$\tilde{x}$ of the Equation (1) we have that
\begin{equation}
 ||x - \tilde{x}||_2\leq 2 \lambda_{2t}^{-1} \varepsilon,  \label{RIPweak-app}
\end{equation}
where $||e||_{2}, ||\tilde{e}||_{2}\leq \varepsilon$. Hence
(10) shows that all solutions of  the Equation (1) are ``almost
equal"  if $\lambda_{2t}$ is large enough. Let us note that for
RS-matrices  any $r$ columns are linear independent
 and hence $\lambda_{2t}>0$, but  $\lambda_{2t}$ tends to zero when $n$ grows and code rate is fixed.
 To find better class of codes over the field of real (or complex) numbers is an open problem.

\section{Conclusion}
\label{conclusion}

In this paper we extends technique, which was developed in
\cite{Lom} for  error correction with errors in both the
channel and syndrome, to the Compressed Sensing problem. We
hope this approach will help to find limits of  the
unrestricted (i.e. without LP usage) compressed sensing.

\end{document}